\documentclass[prd,twocolumn,twoside,nofootinbib,preprintnumbers,superscriptaddress]{revtex4}

\usepackage{amsmath}
\usepackage{amssymb}  % \checkmark
\usepackage{mathrsfs} % \mathscr
\usepackage{graphicx}
\usepackage{dcolumn}
\usepackage{color}
\usepackage{xcolor}
\usepackage{url}
\usepackage{fancybox}

\usepackage[utf8]{inputenc}
\usepackage[T1]{fontenc}

\begin{document}

%\preprint{}

\title{New parameter region in sterile neutrino searches: 
a scenario to alleviate cosmological neutrino mass bound and its
testability at oscillation experiments}

%%%%%%%%%%%%%%%%%%%%%%%%%%%%%%%%%%%%%%%%%%%%%%%%%%%%%%%%%%%%%%%%%%%%%%
\author{Toshihiko Ota}
\email{toshihiko.ota@userena.cl}

\affiliation{Departamento de F\'{i}sica, Facultad de Ciencias, Universidad de La Serena, Avenida Cisternas 1200, La Serena, Chile}
\affiliation{Millennium Institute for Subatomic Physics at the High Energy Frontier (SAPHIR), Fern\'{a}ndez Concha 700, Santiago, Chile}

%%%%%%%%%%%%%%%%%%%%%%%%%%%%%%%%%%%%%%%%%%%%%%%%%%%%%%%%%%%%%%%%%%%%%%
\begin{abstract}
Recent high-precision cosmological data tighten the bound to neutrino
 masses and start rising a tension to the results of lab-experiment measurements, which may hint new physics in the role of neutrinos during the structure formation in the universe.
A scenario with massless sterile neutrinos was proposed to alleviate the cosmological bound and recover the concordance in the measurements of neutrino masses.
We revisit the scenario and discuss its testability at oscillation experiments.
We find that the scenario is viable with a large active-sterile mixing that is testable at oscillation experiments. 
We numerically estimate the sensitivity reach to a sterile neutrino with a mass lighter than active neutrinos in the IceCube atmospheric neutrino observation, for the first time.
IceCube shows a good sensitivity to the active-sterile mixing at the
mass-square difference with a size of $ \sim 0.1$ eV$^{2}$ in the
case of the \textit{inverted-mass-ordering sterile neutrino}, which is
 forbidden under the assumption of the standard cosmology but is allowed
 thanks to the alleviation of the cosmological bound in this scenario.
\end{abstract}

%%%%%%%%%%%%%%%%%%%%%%%%%%%%%%%%%%%%%%%%%%%%%%%%%%%%%%%%%%%%%%%%%%%%%%
\pacs{%
%11.15.Ex % Spontaneous breaking of gauge symmetry
%11.30.Fs %Global symmetries (e.g., baryon number, lepton number)
%11.30.Fs % Flavor symmetry
12.60.-i % Models beyond the SM
%12.60.Cn % Extension of electroweak gauge sector
%12.60.Fr % Extension of electroweak Higgs sector
%12.60.Jv % Supersymmetric models
%13.15.+g % Neutrino interactions
%13.20.He % Decays of bottom mesons
%13.30.-a %Decays of baryons
%13.40.Em % Electric and magnetic moments
%14.20.Dh %Protons and neutrons
%14.40.Nd %Bottom mesons
%14.60.Ef %Muons
%14.60.Pq % Neutrino mass and mixing
14.60.St % Non-standard neutrinos
%14.70.Pw % Other gauge bosons
%14.80.Bn % Standard model Higgs
%14.80.Cp % Non-standard model Higgs
%14.80.Sv % Leptoquarks
%14.80.Mz % Axions and other Nambu-Goldstone bosons (Majorons, familons, etc.)
%23.40.Bw %Weak-interaction and lepton (including neutrino) aspects 
          %(see also 14.60.Pq Neutrino mass and mixing
%95.35.+d % Dark matter
95.55.Vj % Neutrino, muon, pion, and other elementary particle
	 % detectors; cosmic ray detectors 
	 % (see also 29.40.-n Radiation detectors
%96.50.S- % Cosmic rays (see also 94.20.wq Solar radiation and cosmic
          % ray effects) 
%98.80.-k % Cosmology (see also section 04 General relativity and
          % gravitation; for origin and evolution of galaxies, 
          % see 98.62.Ai; for elementary particle and nuclear processes, 
          % see 95.30.Cq; for dark matter, see 95.35.+d; 
	  % for dark energy, see 95.36.+x; for superclusters
	  % and large-scale structure of the Universe, see 98.65.Dx)
98.80.Es % Observational cosmology (including Hubble constant, 
         % distance scale, cosmological constant, early Universe, etc
}


%%%%%%%%%%%%%%%%%%%%%%%%%%%%%%%%%%%%%%%%%%%%%%%%%%%%%%%%%%%%%%%%%%%%%%
\maketitle

%%%%%%%%%%%%%%%%%%%%%%%%%%%%%%%%%%%%%%%%%%%%%%%%%%%%%%%%%%%%%%%%%%%%%%
\section{Introduction}
\label{Sec:Intro}

In coming decades, the standard three-generation neutrino framework will
be extensively tested by the combination of (1) the precision
measurements of $\beta$ decays, (2) the searches for neutrinoless
double beta decays, (3) neutrino oscillation experiments, and (4) cosmological observations.
A brief summary of the current status and future prospects in each
aspect is given in the following;
(1) The precision measurement of the $\beta$ decay spectrum at the
KATRIN detector currently sets the bound $m_{\nu_{e}} \equiv
\sqrt{\sum_{i=1}^{3} m_{i}^{2} |U_{ei}|^{2}} < 0.45$ eV at 90\% CL, and
the bound of the experiment is expected to ultimately reach
$m_{\nu_{e}}<0.2$ eV~\cite{Katrin:2024tvg,KATRIN:2022ayy}, where $m_{i}$
is the masses of neutrinos and $U_{\alpha i}$ $(\alpha \in
\{e,\mu,\tau\})$ is the Pontecorvo-Maki-Nakagawa-Sakata (PMNS) matrix.
For possible future improvements in the $m_{\nu_{e}}$ measurement, cf. e.g., \cite{Project8:2022wqh,Parno:2023upv,Formaggio:2021nfz}.
(2)
No observation of neutrinoless double beta decay processes is pushing down
the upper limit of the effective neutrino mass $m_{ee} \equiv \left|\sum_{i=1}^{3} m_{i} U_{ei}^{2} \right|$.
The KamLAND-Zen collaboration reported $m_{ee} < 0.036-0.156$ eV at 90\% CL~\cite{KamLAND-Zen:2022tow}, and for recent results from the other experiments, see \cite{Majorana:2022udl,CUPID:2022puj,CUORE:2022piu,GERDA:2020xhi,EXO-200:2019rkq}.
The next-generation experiments with various different atoms aim at exploring the entire parameter region of the inverted mass ordering case, cf. e.g.,~\cite{Parno:2023upv,Pompa:2023jxc,Agostini:2022zub,Dolinski:2019nrj}.
(3) The current global fits of the oscillation experiment results show a good
agreement to both the mass orderings~\cite{deSalas:2020pgw,Capozzi:2021fjo,Esteban:2024eli}.
Inclusion of the atmospheric neutrino result by Super-Kamiokande hints the normal ordering but only mildly at $\Delta \chi^{2} = 6.1$~\cite{Esteban:2024eli}.
The two choices in the mass ordering will be determined with a high
confidence level, once the long-baseline accelerator-based oscillation
experiments DUNE~\cite{DUNE:2015lol} and T2HK~\cite{Hyper-Kamiokande:2018ofw}
start running.
The JUNO reactor neutrino oscillation experiment that also aims at
determining the mass ordering is expected to begin to collect data soon~\cite{Stock:2024tmd}. 
(4) Through the parameter fits of the $\Lambda$CDM model extended so as to include neutrino masses, cosmological observations place a tight bound to the sum of the neutrino masses $\sum_{i=1}^{3} m_{i}$.
Using the data of the cosmic microwave background (CMB) and the baryon acoustic oscillations (BAO), the Planck collaboration obtained $\sum m_{i} < 0.12$ eV at 95\% CL~\cite{Planck:2018vyg}.
With the BAO data recently reported by the DESI collaboration, the
combined fit sets the upper limit $\sum m_{i} < 0.072$ (0.113) eV at
95\% CL with a prior of $\sum m_{i}>0 (0.059)$ eV~\cite{DESI:2024mwx,DESI:2024hhd}.
It is expected that new data brought in future by DESI~\cite{DESI:2016fyo}, Euclid~\cite{Amendola:2016saw}, Vera Rubin (formerly known as LSST)~\cite{LSST:2008ijt} 
%,LSSTDarkEnergyScience:2018jkl}, 
et cetera
will improve the bound up to $\sum m_{i} < 0.02$ eV, which covers the entire parameter space of the standard three-generation neutrino framework, cf. e.g.,~\cite{Carbone:2010ik,Hamann:2012fe,Font-Ribera:2013rwa,Basse:2013zua,Brinckmann:2018owf,Chudaykin:2019ock}.
Since those four types of observations provide the independent information on the different combinations of the oscillation parameters, if they will suggest consistently a point in the parameter space, %with a non-zero neutrino mass, 
it could be a highly reliable confirmation of the standard
three-generation neutrino framework which has already been standing the
tests of various oscillation experiments more than two
decades.\footnote{%
For a review of anomalies in neutrino oscillations,
cf. e.g., \cite{Ternes:2023ixc,Acero:2022wqg}.}
However, it is possible that we will encounter a conflict among those observations.
For example, if KATRIN were to find $m_{\nu_{e}} \simeq 0.2$ eV and neutrinoless double beta decays were to be discovered at $m_{ee} \simeq 0.1$ eV, although those observations are consistent with each other in the standard neutrino framework, they would suggest $\sum m_{i} \simeq 0.6$ eV which is already strongly disfavored by the cosmological observations.
For the future prospects of the synergy and possible conflicts between the cosmological observations and the lab experiments in neutrino mass measurements,
cf. e.g., \cite{Gerbino:2022nvz,Gariazzo:2023joe}. 

The cosmological bound is quite powerful, and even it might be possible that the entire parameter region of the standard framework will be 
wiped out.\footnote{%
In fact, the current cosmological observations prefer a \textit{negative
value for the sum of the neutrino masses}, which should be understood as
the enhancement of the clustering in the small scales,
cf. e.g.,~\cite{eBOSS:2020yzd,Craig:2024tky,Green:2024xbb} and
\cite{Naredo-Tuero:2024sgf} for an assessment of the robustness of the
cosmological neutrino mass bound.
Extensions of the standard cosmology to accommodate the feature of the
negative neutrino mass are discussed in e.g.,
\cite{Jiang:2024viw,Elbers:2024sha,Reboucas:2024smm,RoyChoudhury:2024wri}.
Note that it is also pointed out in \cite{Loverde:2024nfi} that a more careful
analysis is needed to claim the preference of the negative mass.
}
%
%While the cosmological bounds can effectively cast the constraints to
%the parameter space of the standard neutrino framework, 
However, it might also require a special care to handle, 
differing from the lab-experiment bounds where
the dependence of the neutrino mass parameters in the signal events is rather simple and the information of the parameters can be extracted from the data in a more clear manner than the case of the cosmological bound.
For the subtleties in the handling of the cosmological bound, cf. e.g.,
\cite{DiValentino:2021imh} and also \cite{Loverde:2024nfi,Bertolez-Martinez:2024wez}.
In addition, the tensions in the $\Lambda$CDM model reported over the last years, cf. e.g., \cite{Abdalla:2022yfr,DiValentino:2021izs,DiValentino:2020vvd}, may suggest that the bound to the sum of the neutrino masses should be discussed and reconsidered with the cosmological models extended so as to solve those tensions.
Nevertheless, it may be worth to consider particle physics models that can
mitigate the cosmological bound in order to understand the currently increasing
stress in the neutrino mass measurements and prepare
for a possible future conflict in the standard neutrino framework.
Central questions that should be addressed in such a study may be
``What can we learn on the physics beyond the neutrino standard model from the conflict (if it will happen)?''
and 
``How can we test such models?''
We will discuss phenomenological aspects of a
model~\cite{Escudero:2022gez}, which is based on the scenario proposed in
\cite{Farzan:2015pca}, and address the questions raised above.
For other attempts to relax the tension in the neutrino mass bounds, 
cf. e.g., \cite{Oldengott:2019lke,Escudero:2020ped,Sen:2024pgb}.

In~\cite{Farzan:2015pca}, Farzan and Hannestad proposed a scenario in which a part of the cosmic neutrino background is replaced with massless fermions. 
%before the structure formation begins.
% 
%Since cosmological observables are only sensitive to the mass of 
%the mixture, the cosmological bound to the mass of neutrinos 
%is weakened.
%
Setting appropriately the mass of a new boson field, which interacts with
both the standard model neutrinos and the massless fermions and promotes this replacement, one can activate the replacement after the big bang nucleosynthesis (BBN) and before the structure formation.
%
%The successful BBN is untouched, because the replacement happens afterwards.
%
The observed acoustic peaks in the CMB spectrum require the decoupled radiation components to be consistent with the standard three-generation neutrinos during the structure formation, cf. e.g.,~\cite{Follin:2015hya,Baumann:2015rya,Baumann:2019keh} and also \cite{Taule:2022jrz}. 
The introduction of the massless fermions in this scenario does not
conflict with this requirement, because the total number of the free-streaming radiation species (=neutrino+massless fermions) is conserved before and after the replacement.
If cosmic history follows this scenario, the current cosmological
observations such as CMB and BAO cast the bound to the masses of the
mixture of neutrinos and the massless fermons, and consequently, the
bound to the actual mass of neutrinos is relaxed.
Notice that since the massless fermions are assumed to reach thermal
equilibrium with neutrinos during the period of the replacement, the
ratio between the number density of neutrinos and that of massless
fermions in the final mixture is determined only by the number (=generations) of the introduced massless fermions.
In short, the bound to the actual sum of the active neutrino masses $\sum
m_{i}|_{\text{real}}$ is simply related to the bound $\sum
m_{i}|_{\text{cosmo}}$ placed by the cosmological observations 
as~\cite{Farzan:2015pca}\footnote{%
The factor of alleviation of the cosmological bound was reestimated in
\cite{Benso:2024qrg} where the thermalization process was handled more carefully.}
\begin{align}
 \sum m_{i} \Big|_{\text{real}}
 =
 \frac{
 1+ \frac{N_{s}}{3}
 }{
 \left(\frac{11}{7} + \frac{N_{s}}{3}\right)^{\frac{1}{4}}
 }
 \sum m_{i} \Big|_{\text{cosmo}}
 \label{eq:bound-mitigated}
\end{align}
where $N_{s}$ is the number of the generations of the massless fermions
(with 2 degrees of freedom, i.e., 2-spinor fields).
The introduction of $N_{s}=23$ massless species mitigates the Planck
bound up to $\sum m_{i}|_{\text{real}} < 0.6$ eV, which can be
consistent with the possible discoveries at lab-experiments of
$m_{\nu_{e}}\simeq 0.2$ eV and $m_{ee}\simeq 0.1$ eV, which was mentioned above.

A realization of this ingenious mechanism was discussed
in~\cite{Escudero:2022gez},\footnote{%
\cite{Benso:2024qrg} for another realization with a dark matter field.} 
in which massless \textit{sterile} neutrinos that interact with a new gauge boson with a mass of keV are introduced.\footnote{%
The word \textit{sterile} may literally mean a particle without any
interaction, but here we use the term for a standard-model-singlet
fermion field with a mixing to neutrinos, which can be a target 
of the searches at various neutrino oscillation experiments. For the
phenomenology of such \textit{sterile} neutrinos, see e.g.,~\cite{Acero:2022wqg,Dasgupta:2021ies}}
After taking various cosmological constraints into consideration, the authors found that the parameter region viable with the Farzan-Hannestad (FH) mechanism was still left available.
In Addendum of \cite{Escudero:2022gez}, the authors reevaluated the
bound to the mixing between active and sterile neutrinos, taking
account of the effect of a thermal potential in the early
universe~\cite{Weldon:1982bn,Dasgupta:2013zpn,Chu:2015ipa}, and found a weaker bound in comparison with the original estimation.
This reassessment opens up a possibility to test the model at oscillation experiments.
Particularly, the observation of atmospheric neutrinos at the IceCube
detector is a unique tool to explore this class of models, because
the Mikheyev-Smirnov-Wolfenstein (MSW) resonance in the oscillation between
active neutrinos and massless sterile neutrinos occurs in the neutrino
channel, which is advantageous in the event statistics point-of-view
thanks to the difference in the detection cross section between 
neutrinos and antineutrinos, 
in contrast to the ordinary search for a sterile neutrino with a mass of
eV, where the MSW resonance occurs in the antineutrino channel,
cf. Sec.~V in \cite{Dentler:2018sju}.\footnote{%
For pioneering works on sterile neutrino searches at IceCube, see
\cite{Nunokawa:2003ep,Choubey:2007ji,Razzaque:2011ab,Barger:2011rc,Razzaque:2012tp,Esmaili:2012nz,Esmaili:2013fva}.}
So far, the IceCube search for a sterile neutrino has been studied only in
the \textit{normal-ordering} case: $m_{\text{active}}^{2} <
m_{\text{sterile}}^{2}$~\cite{Kopp:2013vaa,IceCube:2016rnb,Dentler:2018sju,IceCube:2020phf,IceCube:2020tka,IceCube:2024kel,IceCube:2024uzv},
because sterile neutrinos with the \textit{inverted} mass ordering 
with a size of $|\Delta m^{2}| \sim 0.1-1$ eV$^{2}$, 
where the IceCube experiment has a good sensitivity, 
are largely disfavored by the cosmological bound.
In this model, however, the cosmological bound is alleviated exactly due
to the introduction of the massless sterile neutrinos, and therefore, the
parameter space of the \textit{inverted ordering} case becomes opened up.
We will demonstrate the capability of the IceCube atmospheric neutrino observation in the search for massless sterile neutrinos through a numerical simulation and discuss the testability of this class of models.
Throughout this paper, we use the abbreviations sNO and sIO to indicate
the ordering of the mass-square difference between active and sterile
neutrinos in order to avoid the confusion with the ordinary use of the normal and inverted ordering for the atmospheric mass-square difference;
sNO for the sterile normal-ordering case $m_{\text{active}}^{2} <
m_{\text{sterile}}^{2}$, and sIO for the sterile inverted-ordering case
$m_{\text{active}}^{2} > m_{\text{sterile}}^{2}$.

This paper is organized as follows; 
In Sec.~\ref{Sec:model}, the model to provide the FH mechanism, which
was proposed in \cite{Escudero:2022gez}, will be described.
We will revisit the cosmological bounds to the model parameter space.
In Sec.~\ref{Sec:Lab-bound}, after having a brief discussion on the
constraint to the active-sterile mixing from the global bound to the
non-unitarity of the PMNS matrix, we will present our numerical
estimation of the
sensitivity reach to the active-sterile mixing in the atmospheric
neutrino observation at the IceCube experiment in the sIO case and discuss the testability of the model.
Finally, we will summarize our results and give an outlook on future works in Sec.~\ref{Sec:summary}.

%%%%%%%%%%%%%%%%%%%%%%%%%%%%%%%%%%%%%%%%%%%%%%%%%%%%%%%%%%%%%%%%%%%%%%
\section{Model}
\label{Sec:model}

To realize the FH mechanism, the standard model (SM) must be extended at least so as to contain massless fermions that interact with a mediator field. 
In \cite{Escudero:2022gez}, the authors introduced
four species of new fields in total;
(1) A gauge boson $Z'$ associated with a new $U(1)_{Z'}$ symmetry,\footnote{Although the kinetic mixing term between $U(1)_{Z'}$ and the SM $U(1)_{Y}$ is not forbidden~\cite{Holdom:1985ag} and the $Z$-$Z'$ mixing can arise from the vacuum-polarization-type 1-loop diagram with the active-sterile neutrino mixing, we do not introduce it in our discussion. The term can be controlled by assuming high-energy completions with the unification of gauge symmetries.}
(2) A SM-singlet scalar field $\Phi$ which possesses $+1$ charge of $U(1)_{Z'}$,
(3) $N_{s}$ copies of massless sterile (=SM-singlet) neutrinos
$\nu_{s}$, which are left-handed fermions (2-spinors) and transform
under $U(1)_{Z'}$ with a charge $-1$,
and
(4) 3 generations of heavy right-handed neutrinos $\nu_{R}$ which are singlet under both the SM gauge groups and $U(1)_{Z'}$.
All the SM fields are assumed to be singlet under $U(1)_{Z'}$.
After $\Phi$ takes a vacuum expectation value (vev) $\langle \Phi \rangle = v_{\Phi}/\sqrt{2}$, $Z'$ acquires a mass $M_{Z'} = g_{Z'} v_{\Phi}$ where $g_{Z'}$ is the gauge coupling of $U(1)_{Z'}$.
The spontaneous breaking of $U(1)_{Z'}$ brings a new Higgs boson with a mass around $v_{\Phi}$, which interacts with neutrinos through the active-sterile mixing, and therefore, the coupling is expected to be suppressed. 
We do not discuss the phenomenology of this neutral boson field in the current paper but just mention here that it does not disturb the rest of our discussion.\footnote{%
In \cite{Escudero:2022gez}, $N_{s}$ {\it right-handed} massless sterile
neutrinos were additionally introduced, which only had the $U(1)_{Z'}$
gauge interaction (with a charge $-1$) and did not participate the
neutrino mass matrix. Those additional massless states contribute to the
mitigation of the cosmological neutrino mass bound in the same way as the
left-handed ones, i.e., $\chi$ in \cite{Escudero:2022gez} is understood
as a Dirac fermion with 4 degrees of freedom, which consists of left-
and right-handed massless sterile neutrinos. In that case, $N_{s}$ in
the alleviation formula Eq.~\eqref{eq:bound-mitigated} should be
replaced with $2N_{\chi}$.}

The relevant part of the model Lagrangian is given at Eq.~(3.1) in \cite{Escudero:2022gez}, which is
\begin{align}
 - 
 \mathscr{L}
 \supset&
 \overline{\nu_{R}}
 Y_{\nu}
 L \text{i} \sigma^{2} H
 +
 \frac{1}{2} 
 \overline{\nu_{R}} M_{R} {\nu_{R}}^{c}
 +
 \overline{\nu_{R}}
 Y_{\Phi}
 \nu_{s}
 \Phi
 +
 \text{H.c.},
\end{align}
where $L$ and $H$ are the lepton and Higgs doublets, and $\text{i}\sigma^{2}$ is the anti-symmetric tensor for the $SU(2)_{L}$ doublet representation.
The superscript $c$ denotes the charge conjugation.
According to the number of generations of the fields, the Yukawa couplings $Y_{\nu}$ and $Y_{\Phi}$ are $3\times 3$ and $3\times N_{s}$ matrices respectively.
The Majorana mass $M_{R}$ for $\nu_{R}$ is a $3\times 3$ matrix which is taken to be diagonal with positive real eigenvalues without loss of generality.

After the scalars acquire the vevs, the neutral fermions $\psi_{L} \equiv (\nu_{L}, {\nu_{R}}^{c}, \nu_{s})^{\sf T}$ obtain the mass term
\begin{align}
 \mathscr{L}
 \supset
 -\frac{1}{2}
 \overline{{\psi_{L}}^{c}}
 \mathcal{M}
 \psi_{L}
 +
 \text{H.c.},
\end{align}
where the $(N_{s}+6) \times (N_{s} +6)$
mass matrix is given as
\begin{align}
\mathcal{M}
 =
 \begin{pmatrix}
  0 & m_{D}^{\sf T} & 0
  \\
  m_{D} & M_{R} & \Lambda
  \\
  0 & \Lambda^{\sf T} & 0
 \end{pmatrix}
\label{eq:total-mass-matrix}
\end{align}
with the sub-matrices $m_{D} \equiv v_{\text{EW}} Y_{\nu}/\sqrt{2}$ and $\Lambda \equiv v_{\Phi} Y_{\Phi}/\sqrt{2}$ where $v_{\text{EW}}$ is the vev of the SM scalar field $\langle H \rangle = (0, v_{\text{EW}}/\sqrt{2})^{\sf T}$.
Assuming $\Lambda \ll m_{D} \ll M_{R}$, one can diagonalize this mass
matrix $\mathcal{M}$ with a unitary matrix $\mathcal{U}$, which will be
given below, as
\begin{align}
 \mathcal{U}^{\sf T} \mathcal{M} \mathcal{U}
 =\text{diag}
 \left(
 m_{\text{active}} , m_{\text{heavy}} , m_{\text{sterile}}
 \right),
\end{align} 
where $\text{diag}(m_{\text{active}})$ corresponds to the masses $m_{i=1,2,3}$ of the active neutrinos.
The heavy neutrino masses $m_{\text{heavy}}$ are approximately given with $M_{R i=1,2,3}$ as those in the ordinary type-I seesaw mechanism.
The sterile states are left massless after the diagonalization of this type of mass matrix, i.e., $m_{\text{sterile}} = 0$ which is the $N_{s} \times N_{s}$ null matrix.
Note that the radiative corrections to $m_{\text{sterile}}$ are known to be suppressed~\cite{Escudero:2020ped}, and even after counting them into consideration, those sterile neutrinos can still be approximately handled as the massless states.
The $(N_{s}+6) \times (N_{s} + 6)$ mixing matrix $\mathcal{U}$ is
approximately given as~\cite{Escudero:2022gez}, cf. also
\cite{Chun:1995js,Grimus:2000vj,Barry:2011wb,Zhang:2011vh,Heeck:2012bz,Blennow:2011vn,Ballett:2019cqp},
{\small
\begin{align}
 &\mathcal{U} \simeq
 \begin{pmatrix}
  I_{3} 
  - 
  \frac{1}{2} 
  \left(
   \Theta_{R} \Theta_{R}^{\dagger}
  +
   \Theta_{s} \Theta_{s}^{\dagger}
  \right)
  &
   \Theta_{R}
  &
   \Theta_{s}
  \\
  - \Theta_{R}^{\dagger}
  &
  I_{3} - \frac{1}{2} \Theta_{R}^{\dagger} \Theta_{R} 
  & 
  0
  \\
  -\Theta_{s}^{\dagger}
  &
  0
  &
  I_{N_{s}} - \frac{1}{2} \Theta_{s}^{\dagger} \Theta_{s}
 \end{pmatrix}
 \nonumber
 \\
 &\times
 \begin{pmatrix}
  U_{3\times 3} & 0 & 0
  \\
  0 & I_{3} & 0
  \\
  0 & 0 & I_{N_{s}}
 \end{pmatrix},
\end{align}
}where $I_{N}$ is the $N\times N$ identity matrix
and $U_{3\times 3}$ is a $3\times 3$ unitary matrix that diagonalizes
the sub-matrix of the masses for the active neutrinos as
\begin{align}
 U_{3\times 3}^{\sf T} (-m_{D}^{\sf T} M_{R}^{-1} m_{D}) U_{3\times 3}
 =
 \text{diag}(m_{\text{active}}),
% =
% \text{diag}(m_{i}),
\end{align}
which is similar to the ordinary type-I seesaw mechanism.
The sub-matrices $\Theta_{R}$ and $\Theta_{s}$ in the full mixing matrix
$\mathcal{U}$ can be approximately expressed with the elements of the mass matrix Eq.~\eqref{eq:total-mass-matrix} as
\begin{align}
 \Theta_{R} \simeq m_{D}^{\dagger} M_{R}^{-1}
 \text{ and }
 \Theta_{s} \simeq - m_{D}^{-1} \Lambda.
\end{align}
Notice that the active $3\times 3$ part of $\mathcal{U}$, 
which corresponds to the PMNS matrix and was denoted with the symbol $U$
in Sec.~\ref{Sec:Intro}, is now given as
\begin{align}
 N
 =& 
 \left[
 I_{3}
 - 
 \frac{1}{2} 
 \left(
 \Theta_{R} \Theta_{R}^{\dagger} 
 + 
 \Theta_{s} \Theta_{s}^{\dagger} 
 \right)
 \right] U_{3\times 3},
\end{align}
and it is non-unitary.
%
%and there is a subtle issue in the correspondence between 
%the elements $U_{\alpha i}$ and the measured values 
%of the mixing angles~\cite{Enrique!}.

The new gauge interactions, which drives the FH mechanism, are given as
\begin{align}
 \mathscr{L} 
 \supset&
 \Bigl[
 (\lambda_{Z'}^{\nu_{s} \nu_{s}})_{ij}
 \overline{\nu_{s}}_{i} \gamma^{\rho} \nu_{s j} 
 + 
 (\lambda_{Z'}^{\nu_{L} \nu_{s}})_{ij}
 \overline{\nu_{L}}_{i} \gamma^{\rho} \nu_{s j} 
 \nonumber
 \\
 &
 +
 (\lambda_{Z'}^{\nu_{s} \nu_{L}})_{ij}
 \overline{\nu_{s}}_{i} \gamma^{\rho} \nu_{L j} 
 +
 (\lambda_{Z'}^{\nu_{L} \nu_{L}})_{ij}
 \overline{\nu_{L}}_{i} \gamma^{\rho} \nu_{L j}
 \Bigr]
 Z'_{\rho}
\end{align}
with the couplings
\begin{align}
 (\lambda_{Z'}^{\nu_{s} \nu_{s}})_{ij}
 =& g_{Z'} \delta_{ij},
 \\
 %%%%%
 (\lambda_{Z'}^{\nu_{L} \nu_{s}})_{ij}
 =& 
 - g_{Z'} 
 (N^{\dagger} \Theta_{s})_{i j},
 \\
 %%%%%
 (\lambda_{Z'}^{\nu_{s} \nu_{L}})_{ij}
 =& 
 - g_{Z'}
 (\Theta_{s}^{\dagger} N)_{i j},
 \\
 %%%%%
 (\lambda_{Z'}^{\nu_{L} \nu_{L}})_{ij}
 =& 
 g_{Z'}
 (N^{\dagger}
 \Theta_{s}
 \Theta_{s}^{\dagger}
 N)_{i j}.
\end{align}
The FH mechanism works in this realization as shown in Fig.~2 in
\cite{Escudero:2022gez}, which can be summarized as follows;
Suppose there was no primordial $\nu_{s}$ and $Z'$ at a high temperature.\footnote{%
If $\nu_{R}$s were thermalized in the early universe, their decay would provide $\nu_{s}$ by the era of BBN, which contribute to the extra radiation degree of freedom $\Delta N_{\text{eff}}$.
Therefore, the bound on $\Delta N_{\text{eff}}$ at BBN
can place the extra constraint on $\theta_{s}$ and $N_{s}$. 
However, this constraint can always be circumvented by assuming the reheating temperature lower than $M_{R}$. For more discussion, cf. Sec.~5.1 in \cite{Escudero:2022gez}.} 
When the temperature became $T_{\nu} \sim M_{Z'}$, 
a large part of $\nu_{L}$ took the energy suitable for the resonant pair-annihilation to a $Z'$ through the $\lambda^{\nu_{L} \nu_{L}}_{Z'}$ interaction
and a significant number of $Z'$s were produced and came into
equilibrium with $\nu_{L}$ and also $\nu_{s}$.
Then, when the temperature dropped at $T_{\nu} \lesssim M_{Z'}/3$, $Z'$s
started disappearing from the universe, dominantly decaying to pairs of
$\nu_{s}$s because $|\lambda_{Z'}^{\nu_{s} \nu_{s}}| \gg |\lambda_{Z'}^{\nu_{L} \nu_{s}}| \gg |\lambda_{Z'}^{\nu_{L} \nu_{L}}|$
with small active-sterile mixings
--- The replacement of $\nu_{L}$ with $\nu_{s}$ was completed.

The constraints and requirements to the model parameters were studied at
Sec.~4 in \cite{Escudero:2022gez}, and here we give a brief summary of them.
To activate the mechanism in the period between the BBN and the
structure formation, the value of $M_{Z'}$ must be adjusted around keV.
In \cite{Escudero:2022gez}, 
the authors found that the gauge coupling
should be set at $g_{Z'} \sim 10^{-5} - 10^{-4}$, i.e., $v_{\Phi} \sim
10 - 100$ MeV, to make the mechanism work and simultaneously circumvent 
the constraints from various cosmological observations. 
Interestingly, they also found the lower bound of the active-sterile mixing.
In Ref.~\cite{Escudero:2022gez}, the mixing between active and sterile neutrinos were assumed to be universal, i.e.,
\begin{align}
(\Theta_{s})_{\alpha i} 
=
\theta_{s}
=
\theta_{\nu \chi}
\text{ in Ref.~\cite{Escudero:2022gez}}
\label{eq:universal-theta_s-assumption}
\end{align}
for all the choices of $\alpha=e,\mu,\tau$ and $i=1, \cdots, N_{s}$,
and
the authors showed that the combination of multiple conditions and
bounds shuts the viable parameter region, if the mixing $\theta_{s}$ is
taken to be too small.\footnote{%
Notice that this flavour-universal active-sterile mixing is an
assumption and is not a requirement. To activate the FH mechanism, the mixing between an active flavour and a sterile neutrino is sufficient.}
They concluded that the scenario requires $\theta_{s} \gtrsim 10^{-4}$.
They also evaluated the upper bound to the mixing from the contribution of the
sterile neutrinos to the effective number of neutrinos $N_{\text{eff}}$
at the BBN era.
It is expected that a significant amount of sterile neutrinos could be produced through the oscillation between active and sterile neutrinos, if the mixing is too large. 
However, it is known that the mixing can be strongly suppressed due to
the thermal potential effect in the early universe regardless the
original value of the mixing~\cite{Weldon:1982bn,Dasgupta:2013zpn,Chu:2015ipa}.
The authors of \cite{Escudero:2022gez} took this suppression effect into
consideration and reassessed the upper bound to the active-sterile
mixing in Addendum of their original paper, and they found $\theta_{s} \lesssim
10^{-1}$ is still available, which is accessible at oscillation experiments.
In the left panel of Fig.~\ref{Fig:sinSq2T-DMSq}, 
the bounds and the requirement listed above are summarized
in the parameter plane of the mass $M_{Z'}$ of the gauge boson
and the gauge coupling $g_{Z'}$,
where the active-sterile mixing is assumed to be flavour universal and 
the value is fixed at $\theta_{s}=10^{-2}$, which is within the range of
the allowed values $10^{-4} \lesssim \theta_{s} \lesssim 10^{-1}$.
In the blank region, the cosmological neutrino mass bound is alleviated
with the FH mechanism as Eq.~\eqref{eq:bound-mitigated} and
simultaneously the constraints from the BBN (the excluded parameter
regions are indicated with blue and grey in the plot) and the structure formation (green) are circumvented.

Finally, the Majorana masses $M_{R}$ for the right-handed neutrinos are left for free parameters. 
In this study, we assume that $M_{R}$s are as heavy as the scale of the grand unification a la ordinary seesaw scenario, i.e., $Y_{\nu}$ is set to be roughly order one so as to reproduce correct neutrino masses $m_{i}$. 
This choice of $M_{R}$s makes $\nu_{R}$s be decoupled from the rest of
our discussion, i.e., $\Theta_{R}$ is negligibly small. 
They are not necessarily so heavy, and
it may be interesting to discuss phenomenology of $\nu_{R}$ with a lower $M_{R}$, however, it is beyond the scope of the current study.
For more discussion on phenomenology of $\nu_{R}$ in this model, see Sec.~5.1 in \cite{Escudero:2022gez}.

%%%%%%%%%%%%%%%%%%%%%%%%%%%%%%%%%%%%%%%%%%%%%%%%%%%%%%%%%%%%%%%%%%%%%%
\section{Bounds from oscillation experiments}
\label{Sec:Lab-bound}

Here we discuss the bounds to the active-sterile mixing, 
$(\Theta_{s} \Theta_{s}^{\dagger})_{\alpha\beta}$ and $\theta_{s}$, 
which are placed by oscillation experiments.\footnote{%
For the studies on the impact of sterile neutrinos lighter than (or as
light as) the active neutrinos on the results of oscillation
experiments, cf. e.g.,
\cite{deHolanda:2010am,Bakhti:2013ora,Thakore:2018lgn,deGouvea:2022kma,Chen:2022zts,Chattopadhyay:2022hkw,Chatterjee:2023qyr,Goswami:2024ahm}.
}

%%%%%%%%%%%%%%%%%%%%%%%%%%%%%%%%%%%%%%%%%%%%%%%%%%%%%%%%%%%%%%%%%%%%%%
\subsection{Non-unitarity of the PMNS matrix}

The mixing between neutrinos and extra states in general makes the PMNS
matrix non-unitary, and the experimental constraints to the deviation
from the unitarity have been extensively studied, 
cf. e.g., \cite{Antusch:2006vwa,Antusch:2014woa,Parke:2015goa,Antusch:2016brq,Fernandez-Martinez:2016lgt,Blennow:2016jkn,Forero:2021azc,Blennow:2023mqx}.
The latest bounds on the effective mixing angles $(\Theta_{s} \Theta_{s}^{\dagger})_{\alpha \beta}$ are summarized at Tab.~1 in the Snowmass report~\cite{Arguelles:2022tki}.
Under the flavour universal mixing assumption
Eq.~\eqref{eq:universal-theta_s-assumption}, the effective mixing, which
is constrained by the unitarity bounds, is reduced to 
\begin{align}
 (\Theta_{s} \Theta_{s}^{\dagger})_{\alpha \beta}
 =
 N_{s} \theta_{s}^{2}
\end{align}
for any flavour choice, and the tightest constraint on $\theta_{s}$ is
placed by applying the bound to $(\Theta_{s}
\Theta_{s}^{\dagger})_{\mu \mu} (=2 \alpha_{\mu \mu})$ in
Ref.~\cite{Arguelles:2022tki}, 
which is obtained from sterile neutrino searches in the averaged
oscillation limit (and is not from electroweak precision measurements):
\begin{align}
 \theta_{s} 
 < 1.4 \cdot 10^{-2} \left( \frac{50}{N_{s}} \right)^{1/2}
 \text{ at 90\% CL}.
\label{eq:nonUni-bound}
\end{align}
This shows that the bounds from oscillation experiments are
already reaching the parameter region where the FH mechanism is viable,
$10^{-4} \lesssim \theta_{s} \lesssim 10^{-1}$, 
cf. Addendum of ~\cite{Escudero:2022gez}.
In the right panel of Fig.~\ref{Fig:sinSq2T-DMSq}, we show the bound
placed by the MINOS+ experiment~\cite{MINOS:2017cae} with the orange
curve, which is the main source of the bound to $\alpha_{\mu \mu}$ given
in \cite{Arguelles:2022tki}, cf. also the global
fit~\cite{Forero:2021azc,Blennow:2023mqx}.
The possible improvement of the bound at future long- and short-baseline
oscillation experiments is discussed in e.g., \cite{Machado:2019oxb,DUNE:2020ypp,Acero:2022wqg}.

%%%%%%%%%%%%%%%%%%%%%%%%%%%%%%%%%%%%%%%%%%%%%%%%%%%%%%%%%%%%%%%%%%%%%%
\subsection{Atmospheric neutrinos at IceCube}
\label{Sec:IceCube-bound}

%%%%%%%%%%%%%%%%%%%%%%%%%%%%%%%%%%%%%%%%%%%%%%%%%%%%%%%%%%%%%%%%%%%%%%
\begin{figure*}[t]
 \unitlength=1cm
  \begin{picture}(7,7.5)
   \put(0,0){\includegraphics[width=7cm]{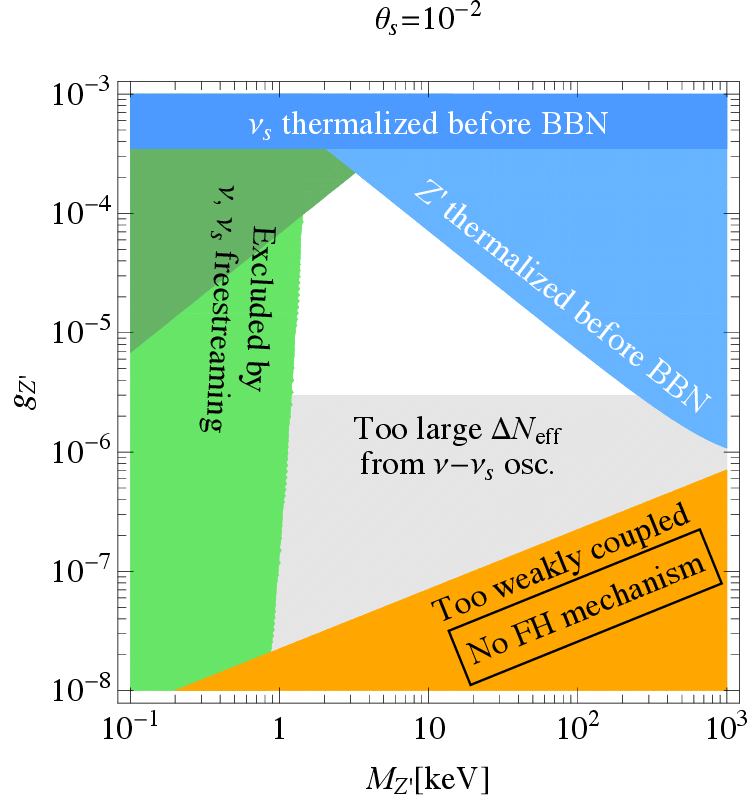}}
  \end{picture}
 \hspace{0.5cm}
  \begin{picture}(7,7.5)
   \put(0,0){\includegraphics[width=7cm]{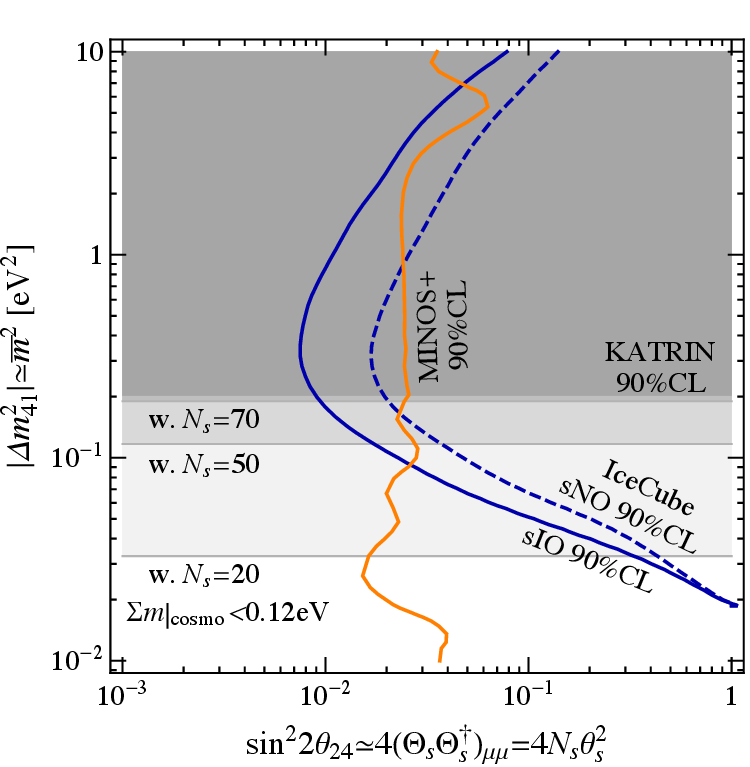}}
%   \put(1.5,6.9){\includegraphics[width=5cm]{plotlegends.eps}}
   \end{picture}
 \caption{%
 [Left]
 Cosmological bounds and the requirement from the FH mechanism in the $M_{Z'}$-$g_{Z'}$ plane, where the active-sterile mixing is fixed at $\theta_{s}=10^{-2}$.
 The blue and grey regions are excluded by the requirements at the BBN era and the green regions are excluded by the observations of the cosmic structures.
 In the orange region, the FH mechanism is not active.
 In short, the FH mechanism works and the cosmological bounds are avoided at the same time in the blank region.
 This parameter region with $m_{\text{active}} \simeq \mathcal{O}(0.1)$ eV can be tested at the IceCube experiment as shown in the right panel.
 [Right]
 Bounds to the absolute value of the mass-square difference $|\Delta
 m_{41}^{2}|$ and the active-sterile mixing $\sin^{2}2\theta_{24}$ from
 the oscillation experiments (curves). 
 %
% The cosmological bound to the square of the averaged mass $\overline{m}^{2}$ of active neutrinos is indicated with shades, which is mitigated due to the introduction of $N_{s}$ massless sterile neutrinos.
 %
 We estimated the sensitivity reach of the IceCube 10.7-year atmospheric
 neutrino data, using the method described in the text.
 Although the sterile inverted ordering (sIO) case (solid blue) is
 only relevant for the FH scenario, we also show the bound in the
 sterile normal ordering
 (sNO) case (dashed blue), which is the ordinary 3+1 model, for comparison.
%, which is in a good agreement with the expected sensitivity reach presented by the IceCube collaboration in \cite{IceCube:2024kel}.
 %
 The grey areas with different shades show the regions excluded by the
 current cosmological bound $\sum m|_{\text{cosmo}}<0.12$ eV 
(by Planck~\cite{Planck:2018vyg}, which is almost the same as
Planck+DESI with the prior $\sum m > 0.059$ eV~\cite{DESI:2024mwx}) 
with the introduction of $N_{s}$ generations of massless sterile
 neutrinos, cf. Eq.~\eqref{eq:bound-mitigated}.
 The darkest grey region is excluded by the measurement of the $\beta$
 decay spectrum at KATRIN~\cite{Katrin:2024tvg}, and 
 neutrinoless double beta decay searches also place a similar bound to KATRIN, 
 cf. text in Sec.~\ref{Sec:IceCube-bound} for more details.
 The introduction of $N_{s} \gtrsim 50$ massless sterile neutrinos opens up the
 parameter region $\Delta m_{41}^{2} = - \overline{m}^{2} \lesssim - 0.1$
 eV$^{2}$ for the sIO case.
 %, and the sterile neutrino search in the IceCube atmospheric neutrino data is the best way to explore that parameter region alleviated by the FH mechanism.
 %
 We also show the parameter region excluded by
 MINOS+~\cite{MINOS:2017cae} with the solid orange curve, which
 approximately corresponds to the bound Eq.~\eqref{eq:nonUni-bound} from the non-unitarity of the PMNS matrix.} 
 \label{Fig:sinSq2T-DMSq}
\end{figure*}
%%%%%%%%%%%%%%%%%%%%%%%%%%%%%%%%%%%%%%%%%%%%%%%%%%%%%%%%%%%%%%%%%%%%%%

It is expected that the search for sterile neutrinos in the IceCube
atmospheric data places a good constraint to the active-sterile mixing
particularly in this scenario, because the relevant mass-square
difference is \textit{inverted ordering} ($m_{\text{sterile}}^{2} =
0 < m_{\text{active}}^{2}$) and therefore, the MSW resonance occurs in
the neutrino channel in differing from the \textit{normal ordering} case
($m_{\text{active}}^{2} < m_{\text{sterile}}^{2}$) in which the MSW resonance occurs in the antineutrino channel.
Thanks to the difference in the detection cross section between
neutrinos and antineutrinos, the search for massless sterile neutrinos
is advantageous in terms of the event statistics over the search for an
ordinary sterile neutrino with a mass of the eV scale, on which the
previous
studies~\cite{Kopp:2013vaa,IceCube:2016rnb,Dentler:2018sju,IceCube:2020phf,IceCube:2020tka,IceCube:2024kel,IceCube:2024uzv}
have consentrated.
The sensitivity to sterile neutrinos with the inverted mass ordering 
has not been studied in the context of the IceCube experiment yet, 
since the introduction of a massless sterile neutrino 
with a mass-square difference of $0.1-1$
eV$^{2}$ suggests 3 active neutrinos with a mass of $0.3-1$ eV, which is
strongly disfavored by the cosmological bound $\sum m_{i}|_{\text{cosmo}}$.
However, in the FH scenario, the cosmological bound is alleviated exactly
due to the introduction of the massless sterile neutrinos, and
therefore, the possibility of the inverted mass-square difference with a size within the range of the IceCube sensitivity is opened up.
Here we present, for the first time, the IceCube sensitivity to sterile
neutrinos with the inverted mass ordering.

In this analysis, we apply the 2-generation fit, in which muon neutrino
(active neutrino with the matter effect only due to the neutral current
interaction) and a generation of sterile neutrino are taken into
account, and therefore, 
there are only 2 parameters: a mass-square difference and a mixing angle.
We call them 
$\Delta m_{41}^{2} \equiv m_{\text{sterile}}^{2} -
m_{\text{active}}^{2}$ and $\theta_{24}$,
following the analysis with the 3+1 model by
the IceCube collaboration~\cite{IceCube:2024kel,IceCube:2024uzv}.
However,
we consider not only the sterile normal ordering (sNO) case 
$\Delta m_{41}^{2} > 0$, which is the 3+1 model,
but also the sterile inverted ordering (sIO) case $\Delta m_{41}^{2} <
0$, which corresponds to the setup of the FH mechanism.
We interpret the bound to the mixing angle $\theta_{24}$ and the mass
square difference $\Delta m_{41}^{2}$ in the 2-generation fit into the bound
to the parameters of the model discussed in Sec.~\ref{Sec:model},
which contains $N_{s}$ massless sterile neutrinos
with the mixing $(\Theta_{s})_{\alpha i}$.
Since the sterile neutrinos are assumed to be degenerate at massless and
the active neutrinos can be also handled as almost degenerate states in
the parameter region to which the IceCube is sensitive, there is only
one relevant mass-square difference, which is the square of the average
of the active neutrino masses $\overline{m}^{2}\simeq m_{i}^{2}$, 
and consequently,
the oscillation formula is greatly simplified and can be directly
compared with the 2-generation oscillation formula. 
In the high-energy region relevant to the atmospheric neutrino
observation at IceCube, the disappearance probability $P_{\mu \mu}$ for
the $\nu_{\mu} \rightarrow \nu_{\mu}$ channel with a constant matter
density is approximately reduced to\footnote{%
The density profile of the Earth's matter is taken into account
in the numerical analysis.
}
\begin{align}
    P_{\mu\mu}
        \simeq&
        1 
        - 
        4
        (\Theta_{s} \Theta_{s}^{\dagger})_{\mu \mu}
        \left[
            \frac{\overline{m}^{2}}{\overline{m}^{2} +  a_{\text{NC}}}
        \right]^{2}
        \sin^{2}
            \frac{(\overline{m}^{2} + a_{\text{NC}}) L}{4E},
\label{eq:Pmumu-reduced}
\end{align}
where $E$ is the energy of a propagating neutrino and $L$ is the baseline length.
The contribution of the matter effect due to the neutral current interaction is given as $a_{\text{NC}} = - \sqrt{2} G_{F} n_{n} E$, where $G_{F}$ is the Fermi constant and $n_{n}$ is the neutron number density along the path of the neutrino propagation.
The formula for the antineutrino channel is obtained by changing the sign in front of $a_{\text{NC}}$ to minus in Eq.~\eqref{eq:Pmumu-reduced}.
Comparing Eq.~\eqref{eq:Pmumu-reduced} to the 2-generation oscillation
formula, cf. e.g., \cite{Choubey:2007ji},
one can find the correspondences between the oscillation parameters in
the 2-generation analysis and the parameters in the model handled in
this study, which are 
\begin{align}
    \sin^{2} \theta_{24}
    =
    (\Theta_{s} \Theta_{s}^{\dagger})_{\mu \mu}
    \text{ and }
    \Delta m_{41}^{2}= -\overline{m}^{2}.
\label{eq:2-gen-params}
\end{align}
Notice again that these relations hold true only in the high-energy region (relevant to the IceCube) and in the small active-sterile mixing limit (which we are interested in).
With the relation Eq.~\eqref{eq:2-gen-params}, we can directly (but
approximately) interpret the result of the 2-generation fit to the bound
to the effective mixing $(\Theta_{s} \Theta_{s}^{\dagger})_{\mu \mu}$ and also the universal mixing angle $\theta_{s}$.

In the right panel of Fig.~\ref{Fig:sinSq2T-DMSq}, we present our numerical
estimation of the expected sensitivity reach with the latest 10.7-year atmospheric neutrino data~\cite{Maltoni2024private}.
The latest data used in \cite{IceCube:2024kel,IceCube:2024uzv} have not
been released to the public yet, and therefore, in order to obtain these
blue curves, we practiced the following method;
First, using the publicly available 1-year data used in \cite{IceCube:2016rnb},  which was published in the data repository,\footnote{%
\texttt{https://icecube.wisc.edu/science/data-releases/}}
we reproduced the exclusion curve presented in \cite{IceCube:2016rnb} 
and calibrated the experimental setups in our analysis.
Next, we replaced the experimental data in our analysis with the
simulated event rates and checked if our sensitivity reach coincided
with the result of the simulation with pseudo-experiments carried out by
the IceCube collaboration, which is indicated with the yellow and green bands 
in Fig.~5 in \cite{IceCube:2016rnb}.
We confirmed that our sensitivity curve ran through the middle of the
green band of the simulation by the IceCube collaboration, i.e., the
experimental setup in our analysis and our simulated event rates are
reasonably consistent with those used in the official IceCube analysis.
Finally, we scaled the size of our simulated event rates to 10.7 years
and obtained the blue curves shown in the right panel of Fig.~\ref{Fig:sinSq2T-DMSq}.
Here we present the sensitivity reaches for both sNO (dashed blue) and sIO
(solid blue) cases, and as we expected from the difference in the event
statistics due to the channels in which the MSW resonance occurs 
(also as mentioned at Sec.~V in \cite{Dentler:2018sju}),
we obtain a better sensitivity in the sIO case.
In the plot, we also show the bound from MINOS+ and find that IceCube
can explore the active-sterile mixing better in the region of $|\Delta m_{41}^{2}| \simeq \mathcal{O}(0.1-1)$ eV$^{2}$.
In the parameter region with $|\Delta m_{41}^{2}| \simeq \overline{m}^{2} =
0.1-1$ eV$^{2}$, the sum of the neutrino masses is 
$\sum m_{i}|_{\text{real}} \simeq 1-3$ eV in the sIO case, 
which is strongly disfavored by the cosmological observations in the standard cosmology.
However, the introduction of $N_{s}\gtrsim 50$ massless sterile
neutrinos alleviates the cosmological bound and opens up the parameter region. 
Notice that the averaged active neutrino mass $\overline{m}$ is also
constrained by the precision measurement of the $\beta$ decay spectrum.
Since the kinetic mass of an electron neutrino is
related to the averaged active neutrino mass as
$m_{\nu_{e}}^{2} = \sum_{i=1}^{3} m_{i}^{2} |U_{e i}|^{2} \simeq
\overline{m}^{2}$
in the
limit where the active neutrino masses are degenerate $m_{i} \simeq
\overline{m}$,
the current KATRIN result~\cite{Katrin:2024tvg} sets the upper bound on
the mass-square difference at
$
 |\Delta m_{41}^{2}|
 \simeq 
 \overline{m}^{2}
 \simeq
 m_{\nu_{e}}^{2}
 <
 0.20
 \text{ eV$^{2}$ at 90\% CL},
$
which corresponds to $\sum m_{i}|_{\text{real}} < 1.4$ eV.
With $N_{s} = 70$ massless sterile neutrinos,
the bound $\sum m_{i}|_{\text{cosmo}} < 0.12$ eV set 
by the Planck collaboration 
is mitigated due to the FH mechanism up to $\sum m_{i}|_{\text{real}} < 1.3$ eV, 
i.e.,
The parameter region opened by the introduction of $N_{s} \gtrsim 70$ is
already excluded by the measurement of the $\beta$ spectrum.
The region excluded by KATRIN is indicated with the darkest grey shade
in the right panel of Fig.~\ref{Fig:sinSq2T-DMSq}. 
%which almost touches the line of the mitigated cosmological bound with $N_{s}=30$. 
%
The neutrinoless double beta decay searches also place the bound to
$|\Delta m_{41}^{2}|$.
The effective neutrino mass is related to the averaged active neutrino
mass as $m_{ee} \simeq \overline{m}\left|\cos^{2} \theta_{12} +
\text{e}^{\text{i} \phi} \sin^{2} \theta_{12} \right|$ in the degenerate
mass limit $m_{i} \simeq \overline{m}$, where $\phi$ is the Majorana CP
phase.\footnote{%
$\phi \equiv \eta_{2} - \eta_{1}$ with the Majorana phases
$\eta_{i=1,2}$ parameterized in the PDG review \textit{Neutrino Masses,
Mixings and Oscillations}~\cite{ParticleDataGroup:2024cfk}.
To obtain the approximated expression of $m_{ee}$, we used the fact
$\sin^{2} \theta_{13} \ll 1$.}
The conservative bound to $|\Delta m_{41}^{2}|$ is obtained by setting
the Majorana phase to $\phi=\pi$ so that the two contributions in $m_{ee}$ are destructive.
In that case, the effective neutrino mass is reduced to $m_{ee} =
\overline{m} \cos 2\theta_{12} \simeq
0.4 \overline{m}$ with the best-fit value of $\theta_{12}\simeq
34^{\circ}$~\cite{Esteban:2024eli}, 
and the bound from KamLAND-Zen~\cite{KamLAND-Zen:2022tow} can be naively interpreted into $|\Delta m_{41}^{2}| \lesssim 8 \cdot 10^{-3} - 0.15$ eV$^{2}$
at 90\% CL with the uncertainty of the nuclear matrix element. 
This shows a similar bound to KATRIN with the choice of the nuclear
matrix element for the lax bound (and also a choice of the value of
$\theta_{12}$ within its uncertainty), 
although the choice for the tight bound leads beyond the degenerate mass regime.

Our analysis~\cite{Maltoni2024private} shows that IceCube has a
sensitivity up to $\sin^{2} 2\theta_{24} \gtrsim 10^{-2}$ (90\% CL) at
$|\Delta m_{41}^{2}| \simeq 0.2$ eV$^{2}$ in the sIO case, 
which is interpreted into the sensitivity to the universal mixing angle 
$\theta_{s} \gtrsim 6-7 \cdot 10^{-3}$ with $N_{s} = 50-70$
massless sterile neutrinos.
In short, IceCube has already started testing the model with the parameters on the blank region presented in the left panel in Fig.~\ref{Fig:sinSq2T-DMSq}, where $\theta_{s}$ is set to be $10^{-2}$, in the case of $\overline{m} \simeq 0.3-0.5$ eV.
Inversely, 
if we will find a sign of sterile neutrinos at IceCube (and/or the other
lab experiment searches for sterile neutrinos), it may suggest the
reconsideration of the cosmological bound in the context of the FH scenario.
If we will have a conflict in the standard three-generation neutrino
framework due to the tight cosmological bound in future, the FH scenario
may be one of the candidates of new physics to explain the situation,
and the search for sterile neutrinos at oscillation
experiments will increase in importance to test the scenario.

Finally, we comment on the necessary number of massless sterile
generations in case where the tight bound $\sum
m_{i}|_{\text{cosmo}}<0.072$ eV suggested by the combined analysis with
the BAO data by DESI (with a prior of $\sum m_{i}>0$ eV)~\cite{DESI:2024mwx} 
is adopted.
To alleviate the bound on the mass-square difference $|\Delta m_{41}^{2}| \simeq \overline{m}^{2} \simeq (\sum m_{i}|_{\text{real}}/3)^{2}$ up to $0.1 - 0.2$ eV$^{2}$, which is the primary region for the IceCube search and is not excluded by KATRIN,
$N_{s} \simeq 90 - 150$ is required.
%
%In short, roughly twice as many massless neutrinos are necessary to
%relax the cosmological bound to $\overline{m}^{2}$ in the same level.

%%%%%%%%%%%%%%%%%%%%%%%%%%%%%%%%%%%%%%%%%%%%%%%%%%%%%%%%%%%%%%%%%%%%%%
\section{Summary and discussions}
\label{Sec:summary}

We revisited the model proposed in \cite{Escudero:2022gez}, which is a
realization of the Farzan-Hannestad (FH) mechanism \cite{Farzan:2015pca} that
alleviates the cosmological bound to the sum of the neutrino masses 
and recovers the concordance with the results of the lab experiments 
in the neutrino mass measurements.
A key ingredient of the model is $N_{s} =\mathcal{O}(10)$ massless
sterile neutrinos with a new abelian gauge interaction.
The bounds from the big bang nucleosynthesis and the structure formation
to the model parameters were discussed in \cite{Escudero:2022gez}.
Recently the bounds were reassessed in Addendum of
\cite{Escudero:2022gez}, and it was found that 
an active-sterile neutrino mixing with the size of 
$\theta_{s} \lesssim \mathcal{O}(0.1)$ was still allowed,
which suggests a possibility to test the model at neutrino oscillation experiments. 
The introduction of sterile neutrinos with an inverted mass ordering 
($m_{\text{sterile}}^{2} < m_{\text{active}}^{2}$) is strongly disfavored by the cosmological bound in the standard cosmology.
However, in this model, the cosmological bound is relaxed with the FH
mechanism, and therefore, a new parameter region --- the sterile
inverted mass-ordering case (sIO) --- is opened up, 
which has not been studied enough in the context of the searches at oscillation experiments. 

In the sterile neutrino search with the atmospheric neutrino data of the
IceCube experiment, the oscillation with massless sterile neutrinos
causes the Mikheyev–Smirnov–Wolfenstein (MSW) enhancement in the
neutrino channel, and therefore, the sensitivity in the sIO case
is expected to be better than that in the search for the ordinary eV sterile neutrinos where the MSW enhancement occurs in the antineutrino channel~\cite{Nunokawa:2003ep,Choubey:2007ji,Razzaque:2011ab,Barger:2011rc,Razzaque:2012tp,Esmaili:2012nz,Esmaili:2013fva}.
We performed a numerical simulation and presented for the first time the
IceCube sensitivity to a sterile neutrino with the sIO mass-square difference~\cite{Maltoni2024private}.
We found that IceCube had already started testing the model with $N_{s} \simeq
50-70$ sterile neutrinos 
in the case with the flavour universal active-sterile mixing $\theta_{s}
\gtrsim 10^{-2}$ and the averaged mass of active neutrinos 
$\overline{m} \simeq 0.3-0.5$ eV.

A comprehensive assessment of the bounds to the active-sterile mixing in
cases beyond the flavour universal assumption is still missing in this study,
which will be discussed in our future work with variations of the model to realize the FH mechanism and phenomenology of the extended neutrino sector and scalar sector.

If the tension in neutrino mass measurements will be heightened in
future, it may indicate new physics in the role of neutrinos in the cosmic history.
The FH mechanism is one of such possibilities, and the searches for
massless sterile neutrinos at various forthcoming oscillation experiments may
provide us a chance to shed light on it.

%%%%%%%%%%%%%%%%%%%%%%%%%%%%%%%%%%%%%%%%%%%%%%%%%%%%%%%%%%%%%%%%%%%%%%
\begin{acknowledgments}

We are grateful to Michele Maltoni for kindly providing his results of the numerical fit of the IceCube atmospheric data. 
We thank Miguel Escudero for clarifying the details of the cosmological bounds on the model and Thomas Schwetz for various discussions, particularly on a new estimation of the relaxation of the cosmological bound, and positive encouragement.
We thank Pilar Coloma and Enrique Fern\'{a}ndez-Mart\'{i}nez for valuable discussions and are grateful to the Instituto de F\'{i}sica Te\'{o}rica UAM/CSIC for warm hospitality.
We also appreciate the members of the SAPHIR TH group for interesting discussions.
We acknowledge support from ANID – Millennium Science Initiative Program
 ICN2019 044.
This work is supported by ANID Chile through FONDECYT grant N${}^{\underline{\text{o}}}$ 1241685.
\end{acknowledgments}

%%%%%%%%%%%%%%%%%%%%%%%%%%%%%%%%%%%%%%%%%%%%%%%%%%%%%%%%%%%%%%%%%%%%%%
% References
%%%%%%%%%%%%%%%%%%%%%%%%%%%%%%%%%%%%%%%%%%%%%%%%%%%%%%%%%%%%%%%%%%%%%%
%\bibliography{./cosmo-mNu-bound}
%\bibliographystyle{apsrev}

\end{document}